\documentclass[pre,aps,twocolumn,superscriptaddress]{revtex4-1}
\usepackage{graphicx}
\usepackage{color}
\usepackage{setspace}
\usepackage{xr}
\usepackage{amsmath}
\usepackage{bm}

\begin{document}
\title{Single-particle level access to hydrodynamic and frictional coupling between spheres in dense colloidal suspensions}

\date{\today}

\author{Taiki Yanagishima}
\affiliation{Department of Chemistry, Physical and Theoretical Chemistry Laboratory, University of Oxford, South Parks Road, OX1 3QZ, United Kingdom}
\author{Yanyan Liu}
\affiliation{Department of Chemistry, Physical and Theoretical Chemistry Laboratory, University of Oxford, South Parks Road, OX1 3QZ, United Kingdom}
\author{Hajime Tanaka}
\email{tanaka@iis.u-tokyo.ac.jp}
\affiliation{Department of Fundamental Engineering, Institute of Industrial Science, The University of Tokyo, Komaba 4-6-1, Meguro-ku, Tokyo 153-8505, Japan}
\author{Roel P. A. Dullens}
\email{roel.dullens@chem.ox.ac.uk}
\affiliation{Department of Chemistry, Physical and Theoretical Chemistry Laboratory, University of Oxford, South Parks Road, OX1 3QZ, United Kingdom}

\begin{abstract}
{\bf
The rotational Brownian motion of colloidal spheres in dense suspensions reflects local hydrodynamics and friction, both key to non-linear rheological phenomena such as shear-thickening and jamming, and transport in crowded environments, including intracellular migration and blood flow. To fully elucidate the role of rotational dynamics experimentally, it is crucial to measure the translational and rotational motion of \emph{all} spheres simultaneously. Here, we develop compositionally uniform colloidal spheres with an off-centre, fully embedded core with a different fluorophore to the particle body, allowing access to rotational motion for all particles at the single-particle level. We reveal interparticle hydrodynamic rotational coupling in charged colloidal crystals. We also find that higher local crystallinity in denser crystals enhances rotational diffusivity, and that nearly arrested particles exhibit a stick-slip rotational motion due to frictional coupling. Our method sheds new light on the largely-unexplored local rotational dynamics of spherical particles in dense colloidal materials.
}
\end{abstract}

\maketitle

The dynamics of colloidal particles is key to connecting the equilibrium phase behaviour of particulate suspensions to their atomic analogues \cite{Poon2004,Philipse2018}. The vast majority of studies of condensed matter phenomena only focus on translational degrees of freedom; examples include crystallisation \cite{Palberg2014,Russo2012,Arai2017}, melting \cite{Wu2009,Wang2012}, gelation \cite{Lu2008,Tateno2019} and the glass transition \cite{Hunter2012} (see Ref.~\cite{Lu2013} for a review). However, this is only half the picture. As noted from the outset by Perrin \cite{Perrin1913}, colloidal particles also feature rotational Brownian motion. This includes spherical particles which do not have an easily visualised orientation. For a suspension of spherical particles, rotational Brownian motion is governed by two physical effects, hydrodynamics and friction. Non-local hydrodynamic interactions play an important role in colloidal gelation \cite{Varga2015,Tateno2019,DeGraaf2019}, the rheology of complex fluids \cite{Stickel2005}, and biological systems \cite{Ando2010,Golestanian2011a,Mourao2014}. Local friction between colloidal particles is also of crucial significance, where surface roughness may directly affect relaxation in dense suspensions \cite{Royer2016}; colloidal ``rolling'' may sometimes even modify phase behaviour \cite{Wang2015}.

Although orientational dynamics have been studied extensively for \emph{anisotropic} particles \cite{Dogic2006b,Muller2013,Smalyukh2018}, studies of rotational fluctuations in dense suspensions of spheres are exceedingly few. This is due to the lack of a colloidal model system that allows both the position and orientation of \emph{all} the spheres in a field of view to be imaged up to arbitrarily high volume fractions. Studies using light scattering have given us valuable insights into rotational diffusion in suspensions \cite{Degiorgio1995,Koenderink2002a,Koenderink2003a}, but have been unable to correlate dynamics with the structure to reflect structural heterogeneity, a key strength of mesoscopic colloidal models.

There has been a clear recent escalation in efforts to create a viable system. Efforts to directly image rotational dynamics include nematic liquid crystal droplets with a frozen director \cite{Reichert2004a}, Janus (MOON) particles \cite{Behrend2005a}, anisotropic fluorescence profiles using photobleaching \cite{Lettinga2004a,Wenzl2013}, and rough colloidosomes with a subpopulation of fluorescent surface probes \cite{Ilhan2020}, none of which allow confocal microscopy studies at arbitrarily high volume fractions. A recent effort encapsulated a silica core \cite{Liu2016} in a 3-trimethoxysilyl propyl methacrylate (TPM) shell to make a core-shell particle, again leaving a scattering interface that makes confocal microscopy in dense suspensions unfeasible. Anisotropic PMMA clusters in a spherical PMMA shell were proposed to rectify this \cite{Schutter2017}, but producing monodisperse batches in bulk is not possible. As of yet, no system allows full three-dimensional (3D) characterisation of monodisperse systems using confocal laser scanning microscopy (CLSM) in arbitrarily dense systems.

Here, we overcome this impasse by developing a bulk synthesis for monodisperse colloidal spheres with uniform composition and non-uniform fluorescence profile which can be density- and index-matched in a solvent mixture, allowing a single 3D confocal microscopy snapshot to reveal the coordinates and orientations of \emph{all} observed particles. Using these probes, we study hydrodynamic coupling in charged colloidal crystals, finding that the rotation of adjacent spheres exhibits transient coupling. Furthermore, by adding a non-fluorescent outer layer, we make particles suitable for studies at arbitrarily high concentrations. We use these to study a dense, partially crystalline sediment, and find, for the first time, a positive correlation between rotational diffusivity and local crystallinity. Finally, we observe a stick-slip dynamics, indicative of the emergence of local contact friction.

\begin{figure*}
\includegraphics[width=17cm]{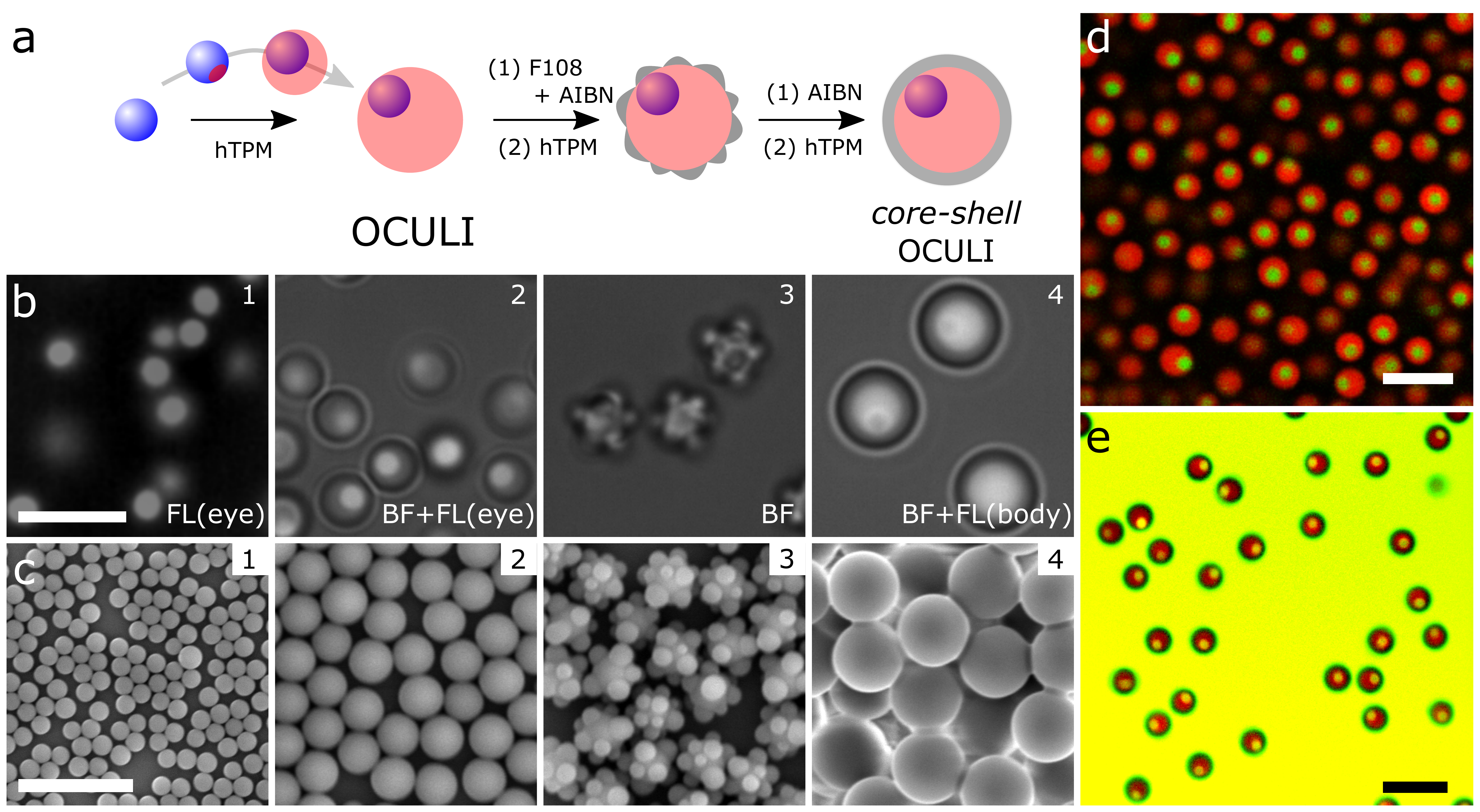}
\caption{{\bf Synthesis of OCULI and core-shell OCULI particles.} (a) Schematic of how monodisperse TPM particles are embedded in pre-hydrolysed TPM (hTPM) to form off-centred `OCULI' particles. Further coating by hTPM in a two-step procedure via intermediate raspberry particles results in `core-shell OCULI particles'. (b1) Fluorescence (FL) image of TPM `eye' particles. (b2) OCULI particles, imaged in bright field (BF) with fluorescence excitation of the eye. (b3) Raspberry OCULI particles, imaged in bright field. (b4) Core-shell OCULI particles, imaged in bright field with fluorescence excitation of the OCULI body. (c) SEM pictures of particles in (b). (d) Two-channel confocal microscopy image of OCULI particles. (e) Two-channel confocal microscopy image of core-shell OCULI particles. The solvent is also dyed with a trace amount of BDP-FL. All scale bars are 5 $\mu$m.}
\label{fig:SYNTH}
\end{figure*}

\section{Colloidal spheres for tracking 3D rotational dynamics}

To facilitate the study of translational and rotational motion of all spheres, our monodisperse, compositionally homogeneous colloidal spheres consist of a core-shell structure with a core localised to the surface in a spherical shell; the core and shell are both TPM, but labelled with different dyes. Given the lack of a name for such spheres, we call these particles \emph{OCULI} particles (`Off-center Core Under Laser Illumination'), and hereafter refer to the core as the `eye', and to the whole particle as the `body'. Importantly, TPM particles can be transferred into a density and index-matching solvent mixture \cite{Liu2016a,Liu2019}, allowing imaging deep inside dense suspensions.

\begin{figure*}
\includegraphics[width=17cm]{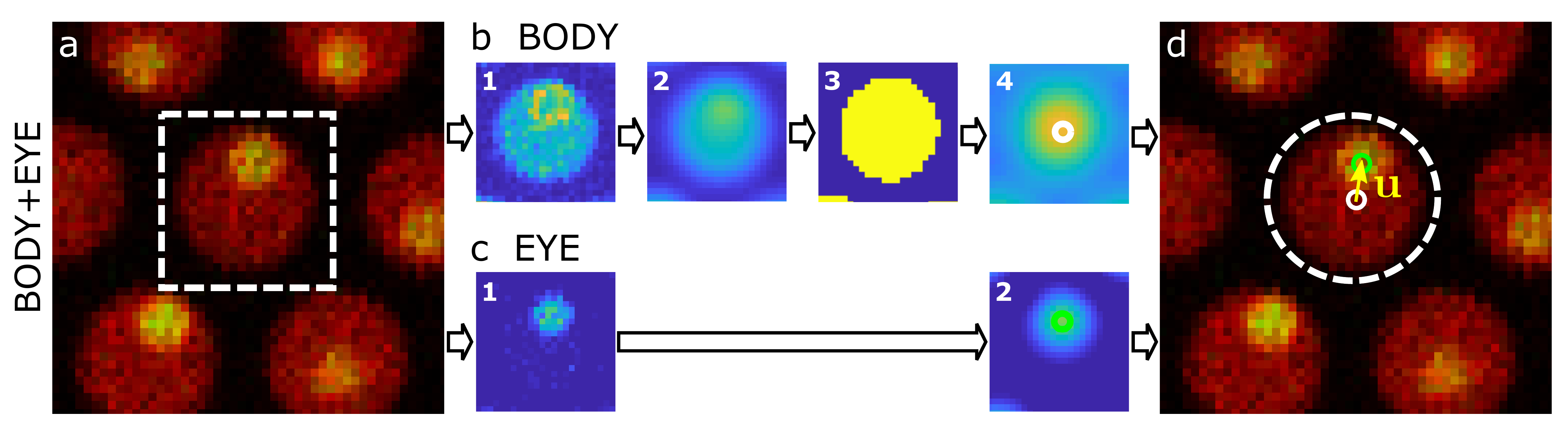}
\caption{{\bf Orientation vector location from confocal microscopy} (a) Composite image of a single slice from a two-channel 3D-CLSM stack. (b) Particle location for the particle body. The raw data for the corresponding channel (1) is band-passed (2), binarised (3), then re-smoothed (4) before a particle location algorithm is applied (point in (4)). (c) Particle location for the particle eye. The raw data for the corresponding channel (1) is band-passed (2) before a particle location algorithm is applied (point in (2)). (d) Body and eye positions are associated via a simple distance threshold (dotted circle) and joined to give a particle specific orientation vector, $\bm{u}$.}
\label{fig:TRACK}
\end{figure*}

\subsection{Synthesis of OCULI particles}
A schematic of the synthesis is given in Fig.~\ref{fig:SYNTH}(a). Monodisperse, fluorescent TPM particles \cite{Liu2016a} are suspended in a basic solution (Fig.~\ref{fig:SYNTH}(b1)). Pre-hydrolysed TPM (hTPM) is added, allowing TPM to condense onto the particles until it engulfs it. Interface pinning \cite{Manoharan2015} and a low wetting angle between hTPM and the core ensures that the core is localised just inside the surface, as shown in Fig.~\ref{fig:SYNTH}(b2). The body is labelled using a spectrally distinct dye and cross-linked. The result is an off-centre core-shell particle that is compositionally isotropic except for a trace amount of dye, critical for realising a spherically symmetric and isotropic particle with no gravitational bias. SEM images (Fig.~\ref{fig:SYNTH}(c1) and (c2)) show that both the core and the OCULI particles are spherical and smooth. A two-channel 3D-CSLM image in an index-matching mixture of trichloroethylene (TCE) and tetralin is also shown in Fig.~\ref{fig:SYNTH}(d). For this study, we choose BDP-FL for the eye and Cyanine3 for the body.

Fully fluorescent particles allow particle localisation when interparticle contact is rare. At higher volume fractions, localisation becomes difficult using 3D-CLSM due to overlapping point spread functions \cite{Leocmach2013}. In the same spirit as symmetric core-shell particles \cite{Dullens2003,Kodger2015}, we add an extra non-fluorescent layer of TPM to the OCULI particles following \cite{Liu2016a}, as illustrated in the latter half of Fig.~\ref{fig:SYNTH}(a). OCULI particles are exposed to Pluronic F108 before cross-linking; adding hTPM now nucleates small surface lobes to make a `raspberry'-like particle (see Fig.~\ref{fig:SYNTH}(b3) and (c3)). These are cross-linked and exposed to more hTPM, filling the gaps and making the particles spherical again. Fig.~\ref{fig:SYNTH}(b4) shows a bright-field image with added fluorescence excitation of the OCULI body inside. A confocal image is also given in Fig.~\ref{fig:SYNTH}(e), suspended in an index-matching solvent mixture with a trace amount of BDP-FL dye, like the eye. A dark, non-fluorescent layer is clearly visible at the surface, while the central part of the particle maintains the unique fluorescence profile of the OCULI particles. SEM again confirms that the core-shell OCULI particle is also spherical and smooth (see Fig.~\ref{fig:SYNTH}(c4)). A detailed protocol is provided as Supplementary Information.

\subsection{Tracking the rotational motion of individual particles}
\label{sct:tracking}

To study the rotational dynamics of each particle using 3D-CLSM, we apply conventional particle tracking methods to obtain eye and body positions separately. An example is given in Fig.~\ref{fig:TRACK}; a two-channel slice of a 3D-CLSM stack is shown in (a). Eyes were tracked using a variant on \cite{Crocker1996}: the eye signal (c1) is smoothed before a Gaussian fit around the maximum gives a position with sub-pixel accuracy (c2). The body is located by taking the body signal (b1), band-passing it (b2) and binarising (b3) before re-smoothing using a Gaussian kernel (b4) to reduce any bias introduced by the eye. Sub-pixel accuracy is achieved with a quadratic fit around the maximum. Eyes are associated with bodies by a simple distance threshold, as shown in Fig.~\ref{fig:TRACK}(d). The vectors joining them are normalised to give a unit orientation vector $\bm{u}(t)$ associated with each particle. Standard methods \cite{Crocker1996} are used to associate body positions into trajectories over time. Particles with multiple eyes ($<0.5$\% of total) are removed from the analysis.

\section{Rotational dynamics of \emph{all} spheres in colloidal materials}

Rotational Brownian motion at high densities is chiefly governed by non-local hydrodynamic and local frictional interactions. To quantitatively address these phenomena and elucidate rotational correlations, it is a key prerequisite to have access to both the position and orientation of \emph{all} the spheres, something which our OCULI particles deliver for the first time in dense materials. We present two relevant case studies: hydrodynamic coupling in charged colloidal crystals and the relation between rotational diffusivity, crystallinity and friction at the single-particle level in dense colloidal sediments. 

\begin{figure*}
\includegraphics[width=1.0\linewidth]{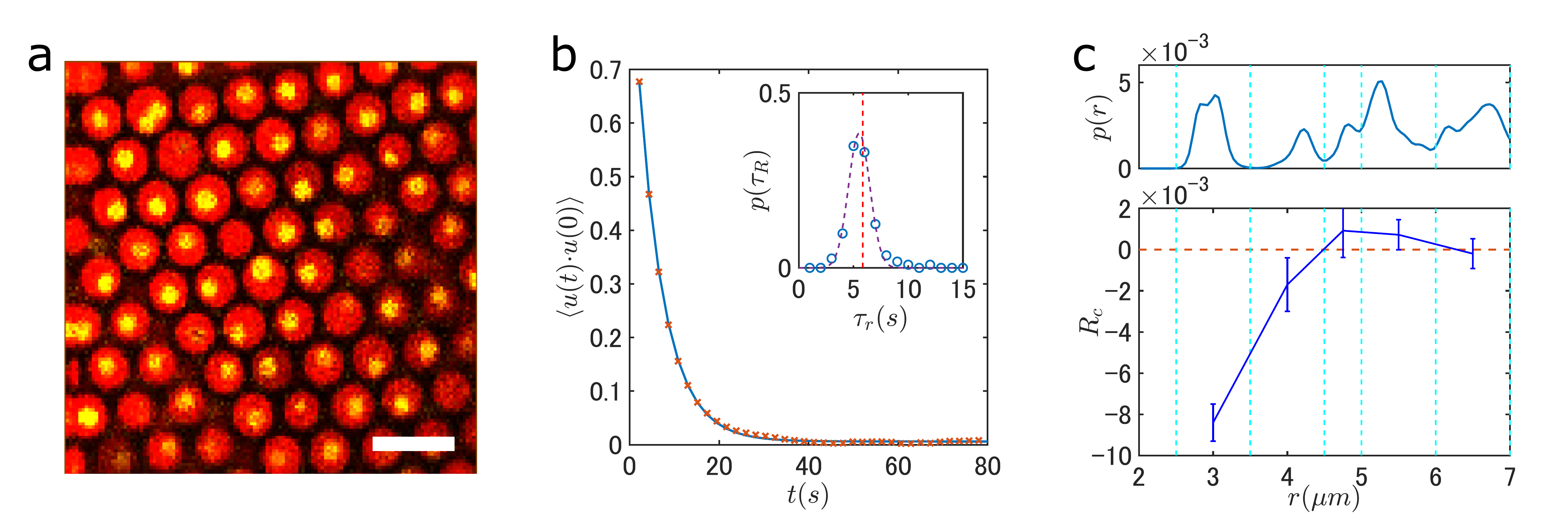}
\caption{{\bf Rotational motion in a charged colloidal crystal of OCULI particles.} (a) Two-channel confocal microscopy image of OCULI particles. (b) Average autocorrelation function of orientation vectors over time. (Inset) Distribution of relaxation times for single particles in a field of view (symbols) and a Gaussian fit (solid line). Red dotted line shows the average relaxation time found from the exponential fit to the average autocorrelation function. (c) (upper) Probability distribution $p(r)$ of inter-particle distances in the charged crystal. (lower) Rotation-rotation coupling constant $R_c$ (Eq.~\ref{eq:C}) for different coordination shells around a central particle (separated by dotted lines) for $\tau_r \approx 2$~s. Error bars correspond to the standard error in the mean value of $R_c$ for particles in each $r$ bin. }
\label{fig:CHARGED}
\end{figure*}

\subsection{Hydrodynamic coupling in charged colloidal Wigner crystals}

When charged particles are not in contact, their rotational motion can couple through hydrodynamic interactions. Colloidal particles in external fields are also known to assemble via hydrodynamically mediated mechanisms, e.g. polystyrene spheres in an alternating field crystallise via collective rotational motion \cite{Santana-Solano2006}. Rotating colloids are even known to possess phases created purely by hydrodynamic interactions \cite{Goto2015,Han2020}. To understand how rotational motion, both passive and driven, underpins assembly and transport, it is crucial to have a complete visualisation of both particle orientations and positions.

To demonstrate the utility of such a measurement, we study rotational Brownian motion in a low-density Wigner crystal of charged OCULI particles. Despite the crowding, the particles are sufficiently separated by electrostatic repulsion to ensure that any rotational correlation is ascribed to hydrodynamic interactions. The absence of long-time translational diffusion also allows for an effective sampling of rotational motion of pairs of particles at certain interparticle separations. A confocal microscopy image is given in Fig.~\ref{fig:CHARGED}(a); the (1 1 1) plane of a face-centred cubic crystal is parallel to the base of the sample cell \cite{Arai2017}. The particles are 1.90~$\mu$m in diameter with a polydispersity of 3.6\% (sized with SEM) suspended in a 1:1 v/v TCE to tetralin mixture with 0.5wt\% OLOA 1200 stabiliser, a concentration low enough to minimise electrostatic screening \cite{Liu2016a}. Though the refractive index of the particles is matched to the solvent to allow 3D-CLSM imaging, the mass density is mismatched, leading to slow sedimentation and subsequent crystallisation. The volume fraction in the crystal is $\sim 0.18$, estimated using the SEM particle size and a 3D Voronoi tessellation. 

We firstly consider single-particle rotational diffusion, calculating the autocorrelation of the orientation vectors $\bm{u}_i(t)$ of each OCULI particle. When a spherical particle undergoes diffusive rotational motion, the autocorrelation function $C(t)$ is expected to decay exponentially,
\begin{equation}
C(t) = \left\langle{\bm{u}(t)\cdot\bm{u}(0)}\right\rangle = e^{-t/\tau_r},
\label{eq:C}
\end{equation}
where $\tau_r = 1/2D_r$, with $D_r$ the rotational diffusion constant \cite{Philipse2018}. At infinite dilution, $D_r$ is given by $D_r = k_{\rm B}T/\pi\eta{\sigma^3}$, where $k_{\rm B}T$ is the thermal energy, $\eta$ is the effective viscosity of the surrounding medium, and $\sigma$ is the diameter of the particle. $C(t)$ averaged over all particles is shown in Fig.~\ref{fig:CHARGED}(b), and is well described by the exponential decay of Eq.~\ref{eq:C}. This shows that the rotational motion of the OCULI spheres, despite their proximity to other particles in the crystal, is purely diffusive at the single-particle level, consistent with previous light scattering experiments \cite{Degiorgio1995}. Given knowledge of individual $C(t)$ for \emph{all} the spheres, we also look at the distribution of relaxation times (Fig.~\ref{fig:CHARGED}(b) (inset)). While the relaxation times from individual particles converge around the average at $\tau_r = 5.8 \pm 0.1$~s, the distribution exhibits a tail towards longer relaxation times. The average relaxation time is also longer than what is expected at this volume fraction \cite{Degiorgio1995}, $\tau_r(\phi = 0.18) = 3.47$~s (see Supplementary Information Section 1D). We attribute this to the gravitational compression of the crystal: the crystal planes are compressed along the vertical direction, making interparticle separations across adjacent (1 1 1) crystal planes smaller than those within the same plane. This is also apparent from the distribution of particle separations $p(r)$ in Fig.~\ref{fig:CHARGED}(c). This compaction should enhance the coupling between the particle size polydispersity and the rotational relaxation. The hydrodynamic friction is significantly stronger for particle pairs of larger size; this may be the origin of the long $\tau_r$ tail in $p(\tau_r)$. Note that  hydrodynamic drag strongly increases with a decrease in the interparticle distance $r$. 

With access to the orientations of neighbouring spheres, we now directly quantify hydrodynamic rotation-rotation coupling between spheres in different coordination shells as identified in $p(r)$ (see Fig.~\ref{fig:CHARGED}(c)). We estimate the angular velocity $\bm{\omega}_i$ of particle $i$ rotating from $\bm{u}_i(t)$ to $\bm{u}_i(t+\tau)$ over a time $\tau$ as $\bm{\omega}_i = \bm{u}_i(t+\tau) \otimes \bm{u}_i(t)$. We thus define a rotation-rotation coupling constant $R_c(r_{ij}=r,\tau)$ given by
\begin{equation}
R_c(r, \tau)  =  \frac{\bm{\omega}_i\cdot \bm{\omega}_j}{\langle \bm{\omega}_i^2 \rangle^{0.5}\langle \bm{\omega}_j^2 \rangle^{0.5}}.
\end{equation}
Note that $R_c$ for pairs of particles will be closer to 1 when they rotate in the same direction, and $-1$ when they rotate in opposite directions. $R_c$ for pairs of particles in the first four coordination shells is shown in Fig.~\ref{fig:CHARGED}(c) for $\tau = 2.17$~s, the time between adjacent frames. We find a weak negative coupling between particles in the first coordination shell, indicating that adjacent particles are more likely to rotate in opposite directions, like meshed gears. Since the first coordination shell is at a distance of $r = 3.0~\mu{\rm m} \approx 1.6 \sigma$, the coupling is mediated by hydrodynamic interactions. Indeed, earlier theoretical work \cite{Reichert2004a} showed that antisymmetric rotation-rotation coupling between Brownian particles due to hydrodynamic effects is non-negligible for this distance. We note that the symmetry of the nearest neighbour particle arrangement in the crystal is incompatible with persistent rotational coupling. As a result, the rotational coupling we have measured is weak and \emph{transient}, and importantly, does not last beyond the nearest neighbours. Importantly, this demonstrates that even weak, transient rotational coupling can be sensitively detected with our OCULI particles.

\section{Rotational diffusivity and local crystallinity in dense, partially crystallised suspensions}

\subsection{Spatial heterogeneity of rotational diffusivity}
\begin{figure*}
\includegraphics[width=1.0\linewidth]{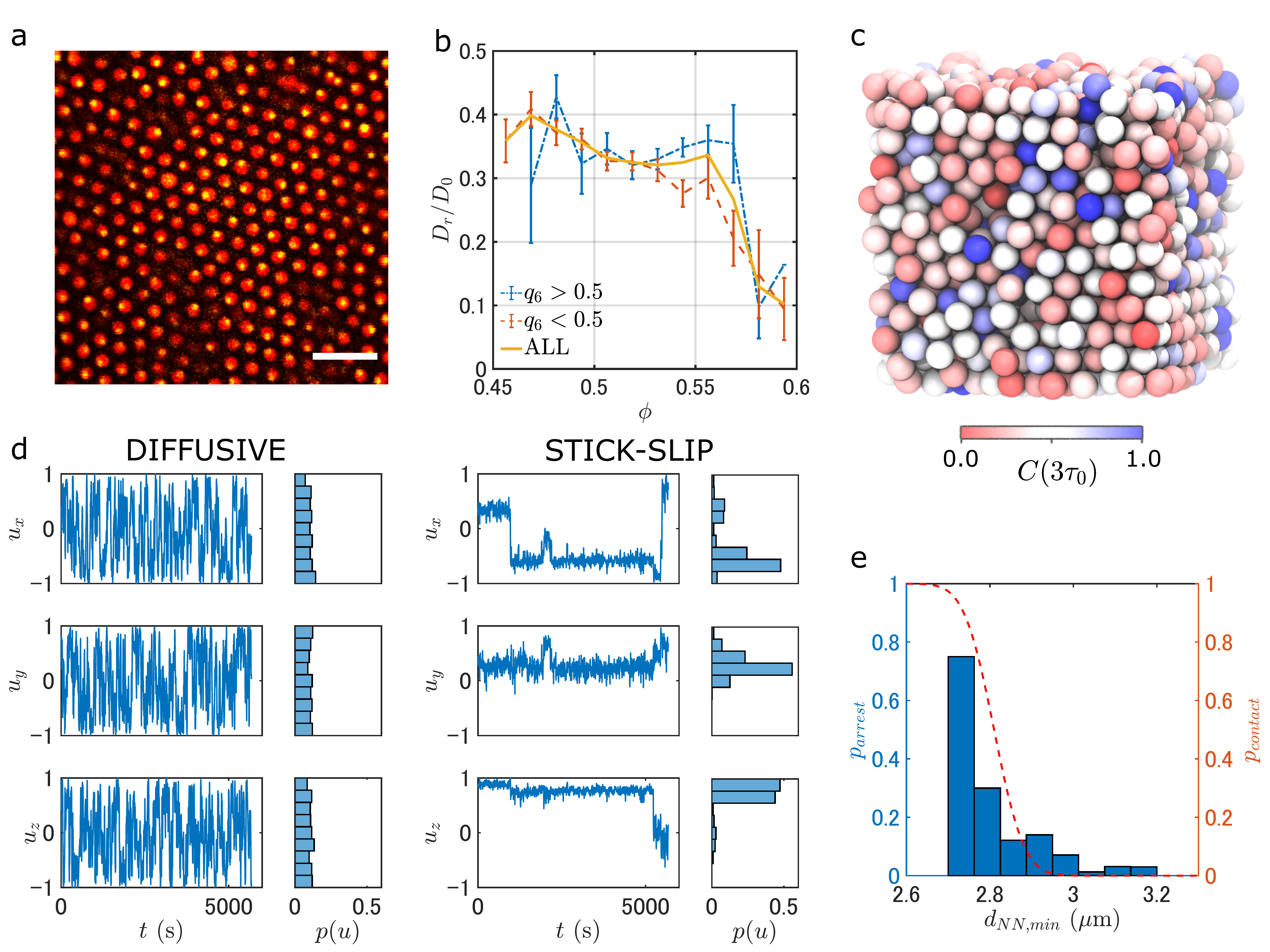}
\caption{{\bf Rotational motion in a partially crystalline sediment of core-shell OCULI particles.} (a) Two-channel confocal image of core-shell OCULI particles. (b) Rotational diffusivity of particles as a function of local volume fraction $\phi$. Statistics are taken over three populations, crystalline particles ($q_6 > 0.5$), amorphous particles ($q_6 < 0.5$) and all particles. (c) A rendering of the particles in the region of interest, colour coded by $C(3\tau_0)$ values: red is for low, blue is for high, scaled over the entire range $0<C<1$. (d) Orientation vector components $u_x,u_y,u_z$ as a function of time for particles undergoing diffusive ($C(3\tau_0) \approx 0.06$, left) and stick-slip  ($C(3\tau_0) > 0.9$, right) rotational motion, and their histograms. (e) Probability of a particle being rotationally arrested, $p_{arrest} = p(C(3\tau_0)>0.9)$, and the probability of two particles being in contact at a particle separation, $p_{contact}$.}
\label{fig:DENSE}
\end{figure*}

Pioneering work on rotational diffusion in dense suspensions \cite{Degiorgio1995,Hagen1999} has measured and simulated how rotational diffusivity varies with volume fraction in density-matched suspensions, with accurate predictions for $D_r$ using a virial expansion. While these works highlight the ultra-sensitivity of $D_r$ to short-range interactions and the pair distribution function, the effect of \emph{local} structure and gravity on local rotational diffusivity remains unaddressed. Access to complete knowledge of $D_r$ of all individual spheres puts us in a unique position to address these ideas in detail. Thus, we form a dense, partially crystalline sediment using core-shell OCULI spheres, as shown in Fig.~\ref{fig:DENSE}(a). The non-fluorescent surface layer ensures a clear separation of fluorescence signals, facilitating accurate particle tracking (see Sec.~\ref{sct:tracking}). The particles have a total diameter of $\sigma = 2.81$~$\mu$m (SEM) with a polydispersity of 2.9\%, and are suspended in a 3:1~v/v mixture of TCE and tetralin. 5\%wt of OLOA 11000 stabiliser is added, keeping the interactions as short-ranged as possible \cite{Sainis2008,Liu2016}. Using bond orientational order parameters, we find that 34.3\% of the sample is crystalline, corresponding to an effective hard-sphere volume fraction of $\sim 51$\% (see Supplementary Information Section 1E).

Firstly, we consider the autocorrelation function $C(t)$ of $\bm{u}(t)$ for each particle (Eq.~\ref{eq:C}) and measure the rotational diffusivity with an exponential fit via Eq.~\ref{eq:C}, using $\tau_r = 1/2D_r$. Hence, we obtain the rotational diffusivity relative to that at infinite dilution $D_r/D_0$ by calculating $D_r / D_0 = \tau_0 / \tau_r$. Here, $\tau_0$ is the rotational relaxation time at infinite dilution, calculated here using the hydrodynamic diameter (see SI). This quantity is plotted as a function of the effective local volume fraction $\phi$ in Fig.~\ref{fig:DENSE}(b). Generally, rotational diffusion slows down upon increasing volume fraction, consistent with earlier light scattering and tracer based experiments \cite{Degiorgio1995,Koenderink2002a,Koenderink2003a,Schutter2017}.

However, there are some key differences with previous work. Firstly, we find that $D_r/D_0$ is smaller than in previous findings \cite{Degiorgio1995,Hagen1999}. Furthermore, we see that at higher local volume fractions, $0.52 < \phi < 0.56$, there is in fact a counterintuitive and surprising plateauing of $D_r/D_0$ with increasing $\phi$ (see solid curve in Fig.~\ref{fig:DENSE}(b)). Previous work found a small levelling-off of translational and rotational diffusivity through the freezing point with increasing $\phi$ \cite{Degiorgio1995}; for translational diffusivity, this was attributed to the increase in free volume upon crystallisation. Note that previous works described the dependence of the average $D_r$ on \emph{average} volume fraction; we now have access to the dependence of $D_r$ of individual particles on their local volume fraction $\phi$. Figure~\ref{fig:DENSE}(b) shows $D_r/D_0$ as a function of $\phi$ for crystalline particles (blue-dashed curve, $q_6>0.5$ and amorphous particles (orange dashed curve, $q_6<0.5$). Strikingly, the crystalline particles exhibit a distinct transient increase in $D_r$, while $D_r$ for the amorphous particles shows a monotonic decrease with $\phi$. Importantly, the two lines clearly lie on either side of the all particle average. To the best of our knowledge, this is the first experimental verification of local crystallinity enhancing rotational diffusion. The fact that an environment with higher order provides less hydrodynamic friction may be rationalised by considering that it is the shortest interparticle separation that mainly contributes to interparticle hydrodynamic friction.

We also directly visualise the rotational heterogeneity in Fig.~\ref{fig:DENSE}(c) by colouring the particles according to their value of $C(t)$ at $t=3\tau_0$. We choose $t=3\tau_0$ to allow for a clear differentiation between slow and fast rotational dynamics: red is $C(3\tau_0) = 0.0$ (fast diffusion, completely decorrelated) and blue is $C(3\tau_0) = 1.0$ (completely arrested). While the heterogeneity is obvious, we also note that there is a subpopulation of particles that are nearly entirely arrested, which will be discussed later.

It remains to address why the rotational diffusivity in our partially crystalline sediment is relatively slow compared to previous studies \cite{Degiorgio1995,Hagen1999} (Fig.~\ref{fig:DENSE}(b)) and remarkably heterogeneous (Fig.~\ref{fig:DENSE}(c)). Firstly, we note that gravity in our system induces significantly more interaction between particles in different layers perpendicular to the gravitational direction, particularly when the charges are screened as they are here. Rotational diffusivity is sensitive to particle proximity and controlled by interaction with the shortest interparticle distance. This effect is more pronounced when the particle arrangement is amorphous. Secondly, particles may even be driven to come into contact with each other, giving rise to a frictional interaction, which is not accounted for in existing theories. Not only would such an effect significantly reduce diffusivity, but it would also amplify the difference between crystalline and amorphous particles, which is exactly what we observe.

\subsection{The emergence of contact friction}
The effect of contact friction may be directly probed using our OCULI particles: this is pivotal to the potential impact these particles may have in other fields. Friction is a key topic for interfacial phenomena at the atomic level \cite{Socoliuc2004,Vanossi2013} and in granular materials \cite{Jaeger1990,Fall2009,Ciamarra2011,Mari2014,Salerno2018} where it may underpin non-linear rheological phenomena such as discontinuous shear thickening \cite{Seto2013,Royer2016}. Some claim that the interparticle frictional profile has a profound effect on suspension behaviour \cite{Comtet2017}. This underlines another key utility of our OCULI system, i.e., how the characterisation of rotational dynamics with single-particle resolution may reveal a spatiotemporal map of frictional forces acting throughout a colloidal packing.

To elucidate the role of contact friction in our partially crystalline sediment, we focus on the subpopulation of particles for which $C(3\tau_0)>0.9$, i.e. are nearly arrested (see dark blue particles in Fig.~\ref{fig:DENSE}(c)). These nearly arrested particles fluctuate in a qualitatively different manner from those with lower $C(3\tau_0)$. Comparing with how the vector $\bm{u}(t) = (u_x,u_y,u_z)$ fluctuates for a particle in the \emph{same sample} undergoing diffusive ($C(3\tau_0) \approx 0.06$) (see Fig.~\ref{fig:DENSE}(d), left), we see that these nearly arrested particles exhibit an intermittent rotational motion (see Fig.~\ref{fig:DENSE}(d), right), with clear peaks in the histogram of $u_x$, $u_y$ and $u_z$. This applies to approximately 70\% of the particles with $C(3\tau_0)>0.9$. Interestingly, this motion is uncorrelated with the translational and rotational motion of its neighbours (see SFig. 3). Instead, it is closely correlated with the distance to the closest nearest neighbour $d_{NN,min}$, as shown in Fig.~\ref{fig:DENSE}(e). The probability of a particle being rotationally arrested $p_{arrest}$ ($C(3\tau_0)>0.9$) rises with shorter $d_{NN,min}$, coinciding with the rise in interparticle contact probability $p_{contact}$, estimated from the Gaussian distribution of particle sizes. 

This result constitutes strong evidence that this peculiar motion arises from contact friction, where the required normal stress is provided by the weight of the particles above it, and that this intermittent dynamics is \emph{stick-slip} motion. This is an essential frictional effect observed on a wide range of length scales \cite{Yoshizawa1993,Palberg1994,Royer2016}. Note that this is a very local effect; one or two larger neighbours or the particle itself being larger might provide enough contact to arrest a neighbour. To the best of our knowledge, this is the first time stick-slip motion has been observed from the rotational Brownian motion of individual colloidal particles.

\section{Conclusions}
We have developed new colloidal model spheres, `OCULI' particles, that enable simultaneous measurement of both the translational and rotational motion of \emph{all} individual particles, regardless of the volume fraction of the suspension. We have applied these to measure transient rotation-rotation coupling in charged colloidal Wigner crystals, a correlation between local crystallinity and rotational diffusivity in a denser, partially crystalline sediment, and the onset of stick-slip friction rotational motion, indicating the emergence of contact friction.

Further to a deeper understanding of self-assembly mechanisms in flowing, driven and non-equilibrium colloidal systems, the ability to directly image variations in the rotational dynamics of individual particles is a significant development for any particulate system where hydrodynamics, mechanical rigidity, and force chain networks play a role. Spatial mapping of rotational diffusivity may thus provide a unique strategy to visualise the incidence of jamming, and allow for accurate characterisation of how the statistics of local force chain networks change on approaching the jamming point. But its greatest potential strength is to change how we scrutinise non-linear rheological phenomena in dense slurries and pastes. Computational work clearly show that contact friction has an important role to play in discontinuous shear thickening \cite{Seto2013,Comtet2017}; strategies are being developed to tune particle roughness to indirectly address the phenomenology \cite{Hsiao2019}. The OCULI system now offers a more direct route, mapping microscopic frictional response to bulk behaviour, an approach which might be applied to any number of complex rheological phenomena at the interface between colloidal and granular matter.

\section{Materials and Methods}
\subsection{Synthesis}
Reagents, preparatory syntheses of core particles, dyes and pre-hydrolysed TPM, as well as a detailed description of the synthesis of both systems described, are given in the SI.

\subsection{Microscopy}
Bright-field microscopy was carried out on an IX71 frame with a 100x PlanApo oil immersion objective (Olympus, Japan). Images were taken with a XIMEA USB3.0 camera MQ042MG-CM (XIMEA Gmbh, Germany). Fluorescence excitation was achieved with a CoolLED Pe-300 white light source (CoolLED, U.K.).

3D confocal laser-scanning microscopy (CLSM) was carried out using the same frame with a 60x PlanApo oil immersion objective (Olympus) and a confocal scan head with a multi-channel laser source (Thorlabs, Germany). 532nm and 488nm lasers were used for two-channel excitation and separated with a standard FITC/Rhodamine filter cube. Scanning in $z$ was achieved using an objective mounted piezoelectric Z-stage (Physik Instrumente, Germany). We used a "Fast-Z" mode, where the piezo element would scan the objective focal plane through the sample, return to its initial position at maximum speed, and repeat. Software and a central control board provided by the manufacturer (Thorsync, Thorlabs, Germany) were used to synchronise the components and record the state of the imaging and piezo at all times to generate accurate timestamps for the frames.

Scan parameters for the two samples are as follows: for the charged colloidal crystal, a 128 x 128 x 251 pixel box was scanned at 2.17~s intervals. Voxels were cubic, with 0.2$\mu$m side length. We chose a spherical region of interest deep inside the interior where the crystal was predominantly monodomain, selecting a cluster of 112 particles to sample over which was well separated from the sides; for the dense crystalline sediment, a 256 x 256 x 411 pixel box was scanned at 5.68s intervals over 1000 frames. Voxel size is the same as above. A region of interest (ROI) is chosen near the base of the colloidal sediment within which we can track 99\% of all particles, excluding particles at the ROI edge. There were $n=$1481 particles to study. Of these, 99.9\% have complete eye trajectories as well.

\subsection{Particle sizing}
All TPM particles are sized using scanning electron microscopy (SEM). The particles are sputter-coated with platinum using a SC7620 sputter coater (Quorum Technologies, U.K.) in an Argon atmosphere, and imaged at 10kV beam energy with a JSM-6010LV scanning electron microscopy unit (JEOL, Japan).

\vspace{5mm}

\bibliographystyle{naturemag}
\bibliography{refclean}

\vspace{5mm}

\noindent
{\bf Acknowledgements}
\noindent
We thank Camille Boulet and Dirk Aarts for assistance with the characterisation of the organic solvent mixtures. T.Y. acknowledges Grants-in-Aid for Young Scientists (B) (JP15K17734) and JSPS Fellows (JP16J06649) from the Japan Society for the Promotion of Science (JSPS); H.T. acknowledges Grants-in-Aid for Scientific Research (A) (JP18H03675) and Specially Promoted Research (JP25000002) from the JSPS; T.Y., Y.L. and R.P.A.D. acknowledge the European Research Council (ERC) (Starting Grant 279541-IMCOLMAT, Consolidator Grant 724834-OMCIDC).

\noindent
{\bf Author Contributions} 
\noindent
T.Y., H.T. and R.P.A.D. conceived the project and designed the experiments. T.Y. performed the syntheses, experiments and analysis. Y.L. prototyped the particles and advised on the experimental work. T.Y. wrote the manuscript; all co-authors discussed and edited the manuscript together. H.T. and R.P.A.D. supervised the project.

\noindent
{\bf Author Information} 
\noindent
Correspondence and requests for materials should be addressed to H.T. (tanaka@iis.u-tokyo.ac.jp) and R.P.A.D. (roel.dullens@chem.ox.ac.uk).

\end{document}


\title{Supplementary Information for `Single-particle level access to hydrodynamic and frictional coupling between spheres in dense colloidal suspensions'}

\author{Taiki Yanagishima}
\affiliation{Department of Chemistry, Physical and Theoretical Chemistry Laboratory, University of Oxford, South Parks Road, OX1 3QZ, United Kingdom}
\author{Yanyan Liu}
\affiliation{Department of Chemistry, Physical and Theoretical Chemistry Laboratory, University of Oxford, South Parks Road, OX1 3QZ, United Kingdom}
\author{Hajime Tanaka}
\email{tanaka@iis.u-tokyo.ac.jp}
\affiliation{Department of Fundamental Engineering, Institute of Industrial Science, The University of Tokyo, Komaba 4-6-1, Meguro-ku, Tokyo 153-8505, Japan}
\author{Roel P. A. Dullens}
\email{roel.dullens@chem.ox.ac.uk}
\affiliation{Department of Chemistry, Physical and Theoretical Chemistry Laboratory, University of Oxford, South Parks Road, OX1 3QZ, United Kingdom}

\maketitle

\section{Supplementary Text}
\subsection{Synthesis of OCULI particles}

\subsubsection{Reagents}
3-trimethoxysilyl propyl methacrylate (TPM) (Sigma Aldrich), aqueous ammonium hydroxide solution (28\%wt solution) (Sigma Aldrich), Cyanine3 NHS Ester (Lumiprobe), BDP-FL NHS Ester (Lumiprobe), anhydrous dimethyl sulfoxide (DMSO) (Sigma Aldrich), 3-aminopropyl trimethoxysilane (APS) (Sigma Aldrich), 4-aminostyrene (Tokyo Chemical Industries), Azobisisobutyronitrile (Sigma Aldrich) and Pluronic F108 (Sigma Aldrich) were used as received. OLOA stabiliser was kindly provided by Eric Dufresne and Azelis. Double distilled water was sourced from a Millipore Direct-Q 3 ultrapure water unit.

\subsubsection{Preparation of fluorescent monomer}
In order to fluorescently label the TPM particles, we require any fluorophore to bear a reactive component which can be incorporated into the TPM network. TPM has two varieties of reactive moieties, three silyl groups and a methacrylate group. Both sites are accessible before the final cross-linking step. We proceeded to label the BDP-FL NHS Ester with APS, and the Cyanine3 NHS Ester with 4-aminostyrene for incorporation into the silyl and methacrylate networks respectively. We note that both dyes can react with both labels. For the BDP-FL, 10 mg of reagent was added to 5 g of chloroform and mixed thoroughly using a PTFE flea and magnetic stirrer. Once the dye was thoroughly dissolved, 10 mg of APS (approx. 5 times molar excess) was added to the mixture, and allowed to react overnight. For the Cyanine 3 NHS Ester, the same procedure was followed using 4-aminostyrene and DMSO. DMSO was not used for the BDP-FL reaction due to adverse effects on the emission profile. After the reaction, both mixtures could be added directly to TPM particle syntheses.

\subsubsection{Synthesis of OCULI particles (charged colloidal crystal)}
We begin by synthesising the `eyes' first. The protocol example here describes how the eyes were made for the eyes of the OCULI particles used to make the charged colloidal crystal; it may be modified following previous works \cite{VanDerWel2017} to change size and yield. 40~ml of double distilled water and 40 ${\rm \mu{l}}$ of 28\%wt ammonium hydroxide solution were stirred in a 100~ml round bottom flask for 15 minutes using an oval PTFE flea. Once a uniform flow was established, 100 ${\rm \mu{l}}$ of TPM was introduced to the flask in one go. This was left to react for approximately 90 minutes. During this time, the TPM hydrolyses and condenses into monodisperse droplets. Next, 100 ${\rm \mu{l}}$ of pre-prepared chloroform/BDP-FL/APS solution (from above) was added to the mixture. Though the chloroform initially collected at the bottom due to its density and immiscibility, the dye gradually diffused into the mixture and transferred to the TPM phase. After 30 minutes of stirring, approximately 50 mg of AIBN was added, and the mixture stirred for 20 minutes. This was transferred to a 50~ml plastic tube and kept in an 80 degree Celsius oven for 3 hours to cross-link the particles. This is subsequently removed and allowed to cool to room temperature. The particles are finally washed at least 5 times using centrifugation (2500RPM for 5 mins per spin) to remove unreacted reagents and disperse the particles in water. The final product is re-suspended in approximately 1.5~ml of double-distilled water. These particles were approximately 1.0-1.2 ${\rm \mu{m}}$ in diameter, and notably monodisperse.

Having obtained a suspension of eye particles, we proceed to coat them with a shell layer to form the body of the OCULI particles. Doing this requires a solution of pre-hydrolyzed TPM (hTPM). Typically, 500 ${\rm \mu{l}}$ of TPM in 5~ml of 0.5~mM hydrochloric acid solution with a PTFE flea. The solution is initially milky but becomes transparent after approx. 1 hour as the TPM hydrolyses.

There are two protocol variations for producing OCULI particles. The first is a stepwise addition of hTPM, recommended for making OCULI particles which are less than 2 microns in size. The introduction of hTPM by hand is fast and straightforward but entails the risk of nucleating new particles instead of growth on the eye. This is less of a problem for fewer addition steps i.e. smaller particles, and was used to produce the OCULI particles in the charged colloidal crystal.

An initial basic solution of eye particles was prepared by taking 0.5~ml of the eye solution and adding 5~ml of 14.8~mM ammonium hydroxide. This is kept in a 20ml glass bottle with a sealed lid. Once the eye particles are well dispersed with a short burst of sonication, 1~ml of the hTPM solution was added, and the bottle quickly tumbled two or three times. Due to the pH, the hTPM is driven to either coalesce or wet onto existing particles. The key here is to ensure that there are enough eye particles in the solution to ensure that wetting onto existing particles occurs at a faster time scale than nucleation, but not so many that multiple eyes will coalesce into single particles.

After 25 minutes, 1.5 ml of hTPM was dripped into the bottle with gentle manual shaking. The solution is left for another 15 minutes. By this point, the pH has dropped due to dilution, so 1 ml of 14.8 mM ammonium hydroxide is added to compensate. We continue to grow the particles by adding another 1.5 ml of hTPM while shaking, like before. This is repeated one more item, and the particles checked using standard optical microscopy.

To dye the particle body, 100 ${\rm \mu{l}}$ of Cyanine3/4-aminostyrene/DMSO dye was added; the tube was then tumbled and left for 20 minutes. Approximately 50 mg of AIBN was then added before the bottle was  tumbled again, left for 20 minutes, then transferred to an oven at 80 degrees Celsius. The particles are allowed to polymerize for 3 hours, with care taken to manually tumble the bottle every hour to prevent particles coalescing at the bottom under gravity. After removal and cooling, they were washed using centrifugation (1200RPM, 5 mins, 5 times) and re-dispersed in distilled water.

\subsubsection{Synthesis of core-shell OCULI particles}
To study dense sediments and jamming, we require particles which are larger and do not suffer the overlap in fluorescence profiles seen in 3D-CSLM with uniformly dyed particles. This requires slightly larger OCULI and the addition of a non-fluorescent layer at the surface. This is based on the protocol described here \cite{Liu2016a}.

The protocol below was used to create the core-shell OCULI used to observe the crystalline sediment. The eyes were produced in a slightly different way; hTPM was pre-hydrolyzed in a 1:10 ratio using a 0.5~mM HCl solution, as above. 5~ml of this hTPM solution was then directly added to 10~ml of 14.8~mM ammonium hydroxide solution in a glass tube before quickly tumbling and leaving stationary. Particles can be seen to nucleate within 20 seconds. The solution is left for 20 minutes to allow all of the hTPM to coalesce into droplets. A small PTFE magnetic flea was added and the mixture stirred slowly ($<$300RPM) to disperse 100$\mu$l of the labelled BDP-FL chloroform solution described above and approximately 10~mg of AIBN. After stirring for 20 minutes, the mixture was allowed to polymerize for 3 hours. These were finally re-dispersed in double distilled water up to a volume of 10~ml. These were kept refrigerated until further use.

Having obtained the eyes, we proceeded to make OCULI particles. Here, we follow an alternative method to demonstrate how the size of OCULI may be tailored to different applications; this is particularly recommended for reaching larger particle sizes ($>2\mu$m in diameter). 5~ml of 14.8~mM ammonium hydroxide was added to a 15~ml plastic tube, followed by 1~ml of the suspension of particles prepared above. The solution was sonicated for 10 minutes to ensure that no clusters become encapsulated instead of single particles. This was followed by 1~ml of a freshly prepared 1:10 hTPM solution before the tube was tumbled and left stationary for 30 mins. This suspension was then added to 20~ml of 14.8~mM ammonium hydroxide solution in a 50~ml round bottom flask with a magnetic flea. The mixture was stirred slowly ($<$100RPM) to ensure homogenisation while preventing the particles from coalescing. 10~ml of hTPM solution was then introduced dropwise over 2 hours using a 10~ml plastic syringe, a PTFE tube and a syringe pump. We note here that the pre-addition of 1~ml of hTPM followed by further dropwise addition led to monodisperse particles, while direct dropwise addition of hTPM to cores did not. This may be due to an energetic barrier to pre-wetting the hTPM which is overcome for different particles at different times given a low concentration of hTPM; this issue is solved by introducing a higher concentration of hTPM initially, followed by the gradual addition of the rest of the material.

If standard OCULI particles are required, 100$\mu$l of dye solution and 20~mg of AIBN should be added to the mixture before stirring for 20 minutes and heating in an 80 degree Celsius oven for 2 hours. However, since the objective of this synthesis was to produce core-shell OCULI, 4~ml of 5\%wt Pluronic F108 solution was added, and the mixture stirred for 20 minutes before adding the dye+AIBN and heating for 2 hours. This is required to modify the wetting properties of the particles for the next step.

Once these OCULI were cross-linked, they were washed thoroughly using centrifugation and double distilled water (1000g spins, 10 mins). Any smaller particles left in the solution at this stage will lead to large secondary particles growing in the following steps. The final particle solution was dispersed thoroughly into double distilled water up to a total volume of 10~ml. Again, this may be refrigerated until further use.

We proceeded to add the non-fluorescent TPM layer. We began by nucleating droplets on the surface of the OCULI to produce `raspberry'-shaped OCULI. 2~ml of the OCULI solution above was added to 100~ml of 14.8~mM ammonia solution in a 250~ml plastic container. 2~ml of freshly prepared hTPM was added before the container was quickly tumbled and left stationary for 45 minutes. 20~mg of AIBN was added before the container was transferred to an 80 degree Celsius oven and left for 3 hours (note that the volume is greater, so the solution requires more time to reach the target temperature). After the added hTPM was cross-linked, the particles were again washed thoroughly with double distilled water using centrifugation (1000g, 10 mins, 5 times) until any particles nucleated in the bulk were removed. The original OCULI are now covered in a dense layer of `lobes', as seen in Fig. 1(b3) and (c3) in the main text. The final suspension was re-dispersed into double distilled water with a total volume of 2~ml, the same volume as the solution of OCULI which was added.

Finally, we filled in the space between the lobes to create smooth, spherical colloidal particles. The 2~ml of `raspberry' OCULI were again added to 100~ml of 14.8~mM ammonia solution in a 250~ml plastic container. This was followed by 2~ml of freshly prepared hTPM solution before the container was quickly tumbled and left stationary for 60 mins. The appearance of the particles were quickly confirmed under bright field microscopy, with particular attention to how circular the diffraction-limited rings surrounding the particles looked. The particles were still rough, so another 2~ml of hTPM was added before tumbling and waiting for another 60 mins. This was repeated as many times as required (in this case, a total of 4 times, including the first addition) to obtain spherical particles. Note that the image will never look perfectly spherical since the refractive index of the cross-linked lobes is different from the uncrosslinked hTPM now residing between the lobes. The best indicator was the shape of the diffraction rings around the particle. Following hTPM addition, 20~mg of AIBN was added before the container was transferred to an 80 degree Celsius oven and heated for 3 hours. The mixture was then recovered, allowed to cool to room temperature before the particles were cleaned again with double distilled water and centrifugation (1000g, 5 mins, 5 times). Finally, the sample was again re-dispersed into 2~ml of double distilled water, ready for use in experiments.

Note that the procedure to transfer these particles from water to organic solvent mixtures is detailed in \cite{Liu2019}.

\subsection{Rotational diffusion in ultra-dilute suspensions}
\label{sec:DILUTE}
For the core-shell system, we considered the diffusion constant of these particles in the ultra-dilute limit; an accurate benchmark was required to quantitatively characterise diffusion constants as a function of local volume fraction. To do this, we modified the density of the solvent slightly, using a 4:1 by volume mixture of TCE and tetralin, to prevent sedimentation and accumulation of the particles at the base of the microscopy cell. We assume the slight change in the composition of the solvent does not adversely affect the particle size except for a change in shear viscosity (1.43 mPas). Sampling trajectories away from surfaces, the autocorrelation of the particle orientations is shown in SFig.~\ref{fig:DILUTE}(a), yielding a relaxation time of $\tau_r = 16.9\pm 0.2$~s. This may be converted to a diffusivity and then an effective hydrodynamic diameter using the equation $D_r = k_{\rm B}T/\pi\eta{\sigma_H^3}$; we find that $\sigma_H = 3.14$ $\mu$m. This is consistent with the size found by SEM given, noting that both particle shrinkage due to drying for SEM imaging and swelling in the haloalkane solvent contribute to making the SEM value smaller. This also allows us to caclaulte an expansion coefficient linking SEM size to hydrodynamic diameter of 10\%.

\subsection{Calculation of coupling constant}
In order to characterise the coupling between the rotation of different pairs of particles, we calculate a coupling constant $R_C$ from an estimate of the angular velocity $\omega$ over time. $\omega$ is estimated from how the orientation vector changes from $\bf{u}(t)$ to time $\bf{u}(t+\Delta{t})$, where $\Delta{t}$ is the time between consecutive frames. $R_C$ may thus be expressed using $u(t)$ as follows:
\begin{eqnarray}
R_c(\tau,r) & = & \frac{\omega_i\cdot \omega_j}{\langle|\omega_i^2|\rangle^{0.5}\langle| \omega_j^2|\rangle^{0.5}}\\
& \approx & \frac{\left\langle{u_i(r_{ij},\tau)\times{u_i(r_{ij},0)}}\cdot{u_j(r_{ij},\tau)\times{u_j(r_{ij},0)}}\right\rangle}{\left\langle{|}{u_i(r_{ij},\tau)\times{u_i(r_{ij},0)}}|^2\right\rangle^{0.5}\left\langle{|}{u_j(r_{ij},\tau)\times{u_j(r_{ij},0)}}|^2\right\rangle^{0.5}} .
\end{eqnarray}

\subsection{Comparison with density matched scenarios for charged crystals}
\label{sec:GRAV}
The single particle rotational diffusivity seen in the charged crystal is slower than what is expected in the ultra-dilute limit using the particle size found from SEM, $\tau_0 = 1/2D_0 = 2.26$~s. This cannot be simply accounted for by swelling of the particle; using an experimentally determined expansion ratio from SEM sizes to hydrodynamic diameters (see Supplementary Text Section \ref{sec:DILUTE}), $\tau_0 = 3.01$~s. Previous work \cite{Clercx1992,Degiorgio1995,Schutter2017} has determined analytical forms for how rotational diffusivity changes with volume fraction. Adopting the expression used here \cite{Degiorgio1995} (also tested against numerical simulations in \cite{Hagen1999}), $D_r = D_0 (1-0.63\phi-0.67\phi^2)$, $\tau_r = 3.47$~s. This is still shorter than the  $\tau_r = 5.8 \pm 0.1$~s found in our experiment.

\subsection{Static structure in a crystalline sediment of core-shell OCULI}
\label{sec:DENSESTATIC}
\begin{figure}[h]
\includegraphics[width=0.8\columnwidth]{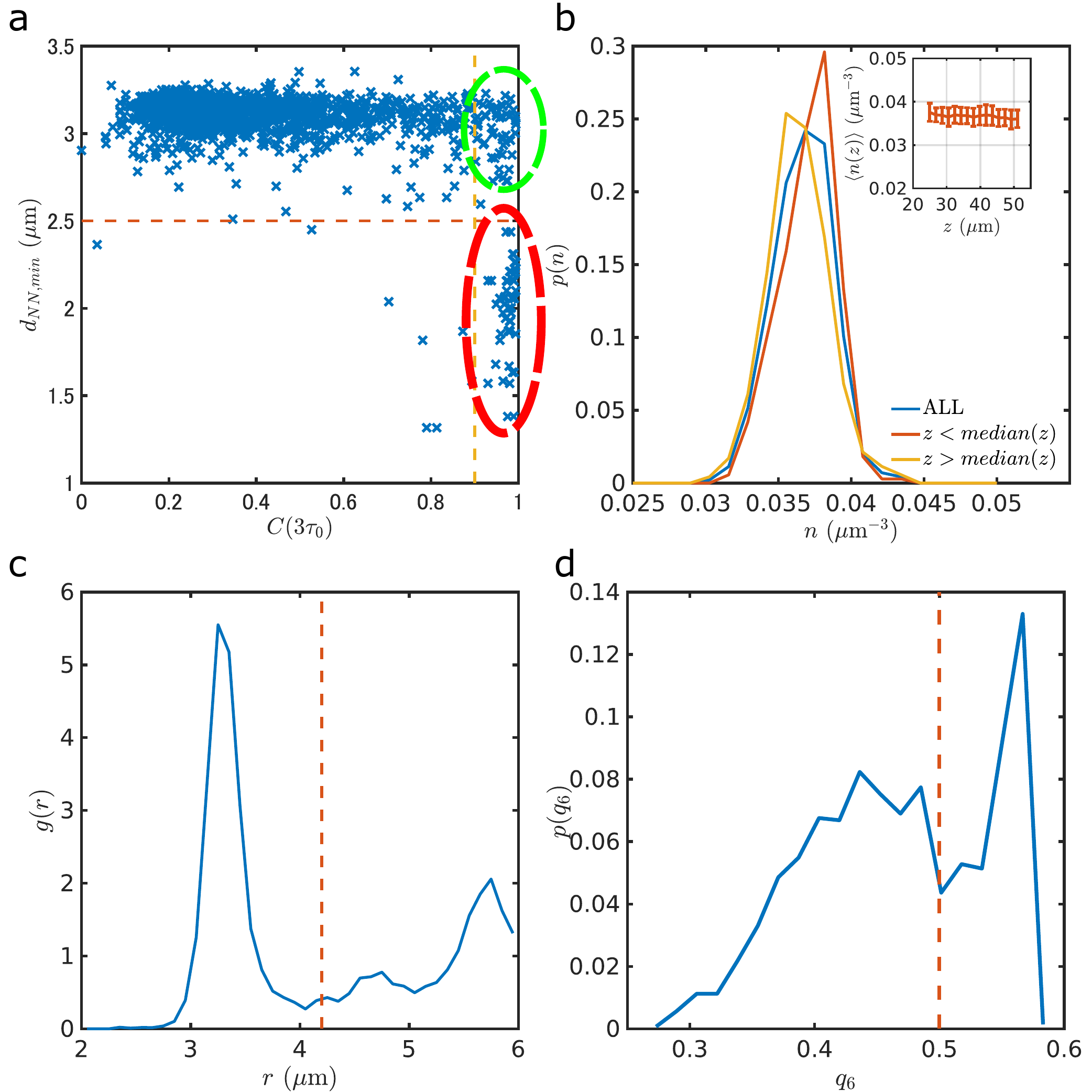}
\caption{{\bf Static properties of dense crystal of core-shell OCULI particles} (a) Minimum distance to neighbouring particles as a function of $C(3\tau_0)$. Particles arrested by frictional interactions are indicated by the upper dotted circle. Particles arrested due to dimerisation with other particles during the growth of the shell are shown by the lower dotted circle. These particles are omitted from the analysis. (b) Histogram of local volume fraction $\phi$ for particles below and above the median $z$ position, as well as the distribution for all particles in the ROI. There is a small variation, but both populations contain a wide range of $\phi$. (inset) Average $\phi$ as a function of $z$. There is a gentle increase with height, as expected in a sediment, but the variation is very small. (c) Radial distribution function. The threshold used for detecting nearest neighbours is given as a dotted line. (d) Distribution of $q_6$ over the whole ROI. The threshold used to separate crystalline and non-crystalline particles is given as a dotted line.}
\label{fig:STATIC}
\end{figure}

\begin{figure}[h]
\includegraphics[width=0.6\columnwidth]{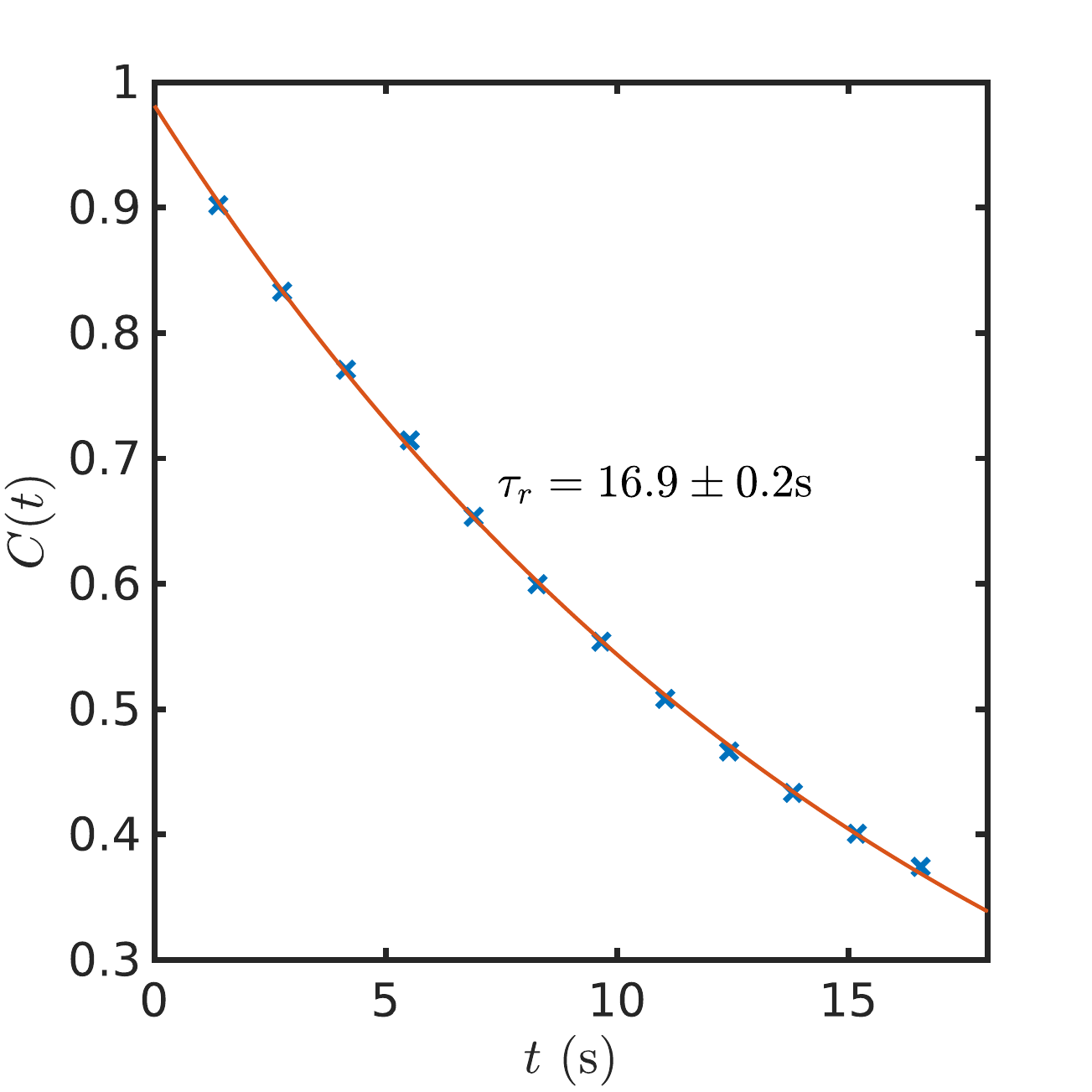}
\caption{{\bf Orientation autocorrelation $C(t)$ for core-shell OCULI in ultra-dilute limit} $C(t)$ for the same core-shell OCULI particles used for the dense crystalline sediment. The data fits well with an exponential decay.}
\label{fig:DILUTE}
\end{figure}

Here, we describe the static structure of the dense crystalline sediment. We firstly noted that there was a small fraction of particles which had fused together during the synthesis, when raspberry particles are formed. These are easily separable by considering minimum distance to a neighbour and the arrest of rotational motion, as shown in SFig. \ref{fig:STATIC}(a). Particles where $d_{NN} < 2.5$ and $C(3\tau_0) > 0.9$ are excluded from all analysis. We also consider broad variation in density over the sample. A Voronoi tessellation allows us to calculate the local number density $n$ for all particles, averaging over 100 frames to lessen the effect of thermal fluctuations. A profile of $n$ with height $z$ is given as an inset in SFig. \ref{fig:STATIC}(b). Note that there is a small reduction with height, but this is very small; this is in agreement with theoretical predictions for the density profile away from the top sediment interface \cite{Dhont1996}. We also see that $n$ for particles above and below the median height have similar distributions (see SFig. \ref{fig:STATIC}(b)), with a similarly wide range.

The system is partially crystalline. We may characterise this by considering the bond-orientational order $q_6$ \cite{Steinhardt1983}, widely accepted as an appropriate local order parameter for crystallisation in hard-sphere like systems \cite{Pusey2009,Tanaka2012}. To associate nearest neighbours, we use a simple distance threshold taken from the first minimum of the radial distribution function $g(r)$ in SFig. \ref{fig:STATIC}(c). With this, we consider the distribution of $q_6$, shown in SFig. \ref{fig:STATIC}(d). One can clearly see that a boundary $q_6\approx 0.5$ separates crystalline and non-crystalline parts of the sample. We may use this threshold and the Voronoi tessellation above to find the proportion of the sample which is crystalline. The sum of the volume associated with particles with $q_6>0.5$ is 34.3\% of the total volume of the ROI. Assuming that the sample is monodisperse, we may use this and the average number density to calculate an effective diameter for the particles,
\begin{equation}
\sigma_{HS} = \left(\frac{6}{\phi_{HS}\pi n_{tot}}\right)^\frac{1}{3} .
\end{equation}
For the sample shown, $\sigma_{HS} = 3.00$ $\mu$m. We use this and local volumes found from Voronoi tessellation to calculate local volume fractions $\phi_i$ in the main text of the paper.


\bibliographystyle{naturemag}
\bibliography{refclean}

\newpage
\section{Supplementary Figures}

\newpage

\newpage
\begin{figure}[h]
\includegraphics[width=0.6\columnwidth]{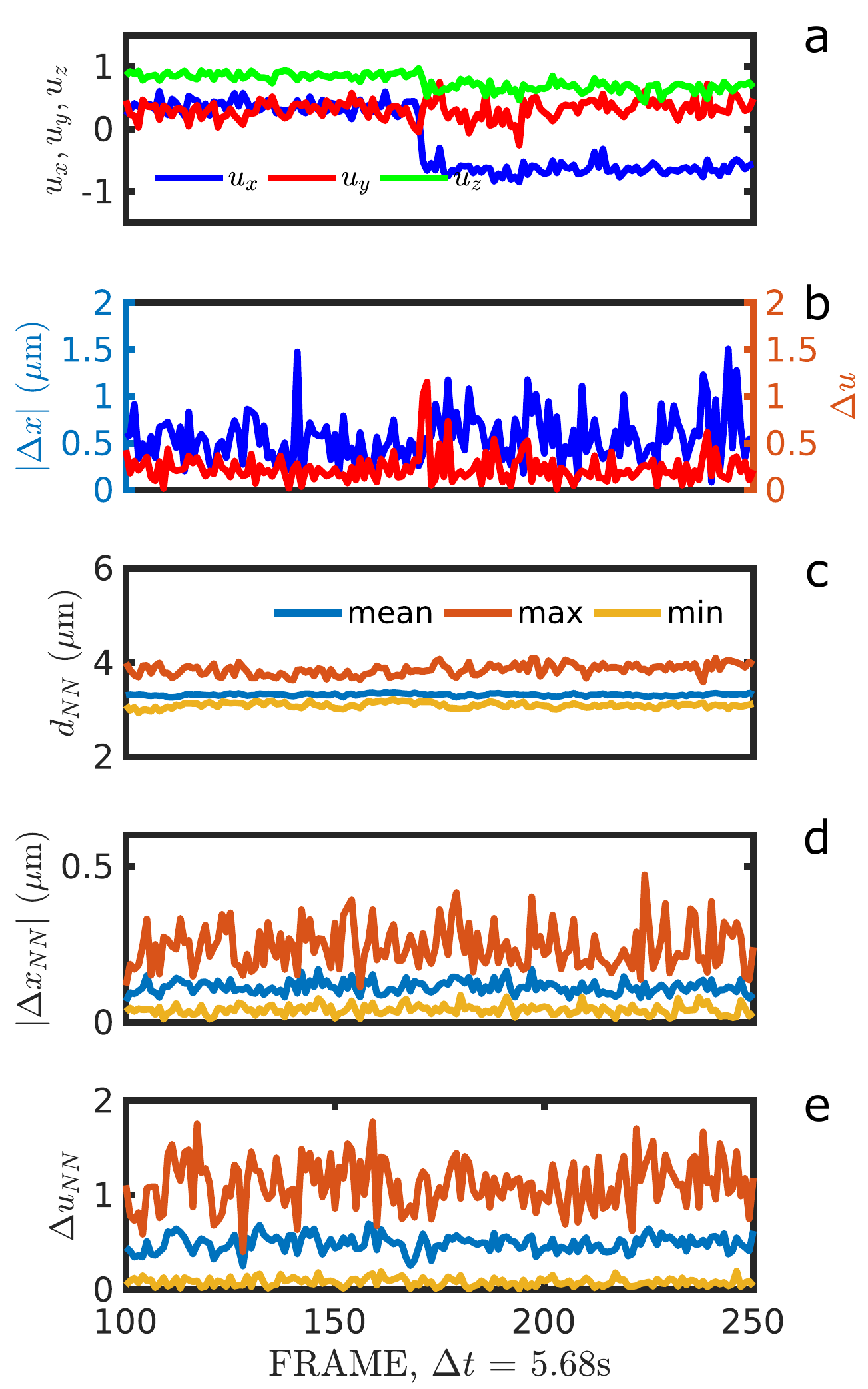}
\caption{{\bf Intermittent dynamics and neighboring particles} (a) $x$, $y$ and $z$ components of the orientation vector of a particle undergoing stick-slip rotational motion as a function of time. All data below refers to same trajectory. (b) Movement of the same particle over two frames, both translational $\Delta{x}$ and orientational $\Delta{u}$ displacements. (c) Distance to nearest neighbours $d_{NN} $as a function of time. (d) Displacements of nearest neighbours over 2 frames as a function of time. (e) Displacements in orientation $\Delta{u}$ over 2 frames as a function of time. (c) to (d) show the mean, maximum and minimum for each statistic.}
\label{fig:STICKSLIP}
\end{figure}

\section{Supplementary Videos}

\subsection{Supplementary Video 1}
Video capture of OCULI particles using confocal microscopy and two laser channels. Both translational and rotational Brownian motion can be seen.
